\documentclass[sigconf, nonacm]{acmart}

 \AtBeginDocument{%
  \providecommand\BibTeX{{%
    \normalfont B\kern-0.5em{\scshape i\kern-0.25em b}\kern-0.8em\TeX}}}

\usepackage{graphics}      
\usepackage{xcolor}
\usepackage{booktabs}
\usepackage{multirow}
\usepackage{upquote}
\usepackage{csquotes}
\renewcommand{\mkbegdispquote}[2]{\itshape}
\AtBeginEnvironment{quote}{\itshape}
\usepackage{tikz}
\usetikzlibrary{decorations.text,calc,arrows.meta}
\usepackage{smartdiagram}
\usepackage[font=small]{caption}
\usepackage{subcaption}
\usepackage{scalerel,graphicx,xparse}

\usepackage{hyperref}
\usepackage[capitalise, nameinlink]{cleveref}

\usepackage{todonotes}
\usepackage{listings}
\usepackage{rotating}

\definecolor{seabornBlue}{RGB}{76,114,176}
\definecolor{seabornGreen}{RGB}{85,168,104}
\definecolor{seabornRed}{RGB}{196,78,82}
\definecolor{seabornOrange}{RGB}{221, 132, 82}

\definecolor{infra}{RGB}{1, 115, 178}
\definecolor{component}{RGB}{222, 143, 5}
\definecolor{pipeline}{RGB}{2, 158, 115}
\definecolor{run}{RGB}{213, 94, 0}

\definecolor{datacoll}{RGB}{204, 120, 188}
\definecolor{expt}{RGB}{202, 145, 97}
\definecolor{evaldeploy}{RGB}{251, 175, 228}
\definecolor{monitorresp}{RGB}{148, 148, 148}

\colorlet{punct}{red!60!black}
\definecolor{background}{HTML}{EEEEEE}
\definecolor{delim}{RGB}{20,105,176}
\colorlet{numb}{magenta!60!black}

\newcommand{\topic}[1]{\vspace{-3.5pt}\smallskip \smallskip \noindent{\bf #1.}}

\def\plaintitle{Operationalizing Machine Learning: An Interview Study}

\makeatletter
\def\url@leostyle{%
  \@ifundefined{selectfont}{
    \def\UrlFont{\sf}
  }{
    \def\UrlFont{\small\bf\ttfamily}
  }}
\makeatother
\definecolor{linkColor}{RGB}{6,125,233}
\definecolor{linkpurple}{HTML}{902043}
\hypersetup{%
  pdftitle={\plaintitle},
  pdfdisplaydoctitle=true, 
  bookmarksnumbered,
  pdfstartview={FitH},
  colorlinks=true,
  citecolor=blue,
  linkcolor=linkpurple,
  urlcolor={blue!80!green!40!black},
  breaklinks=true,
  hypertexnames=false
}

\begin{document}

\title{\plaintitle}

\author{%
Shreya Shankar$^{*}$, Rolando Garcia$^{*}$, Joseph M. Hellerstein, Aditya G. Parameswaran
}
\affiliation{%
\institution{%
University of California, Berkeley \\
{\string{shreyashankar,rogarcia,hellerstein,adityagp\string}}@berkeley.edu \\
{\small $^*$Co-first authors} 
}
\country{}}

\begin{abstract}
Organizations rely on machine learning engineers (MLEs) to operationalize ML, i.e., deploy and maintain ML pipelines in production. The process of operationalizing ML, or MLOps, consists of a continual loop of (i) data collection and labeling, (ii) experimentation to improve ML performance, (iii) evaluation throughout a multi-staged deployment process, and (iv) monitoring of performance drops in production. When considered together, these responsibilities seem staggering---how does anyone do MLOps, what are the unaddressed challenges, and what are the implications for tool builders? 

We conducted semi-structured ethnographic interviews with 18 MLEs working across many applications, including chatbots, autonomous vehicles, and finance. 
Our interviews expose three variables that govern success for a production ML deployment: Velocity, Validation, and Versioning. We summarize common practices for successful ML experimentation, deployment, and sustaining production performance. Finally, we discuss interviewees' pain points and anti-patterns, with implications for tool design.
\end{abstract}

\maketitle








\vspace{-8pt}
\section{Introduction}
\label{sec:intro}

As Machine Learning (ML) models are increasingly 
incorporated into software,
a nascent sub-field called {\em MLOps} (short for ML Operations) has emerged 
to organize the 
``set of practices that aim to deploy and maintain ML 
models in production reliably and efficiently''~\cite{enwiki:1109828739,alla2021mlops}.
It is widely agreed that MLOps is hard.
Anecdotal reports claim that 90\% of ML models
don't make it to production~\cite{weiner2020ai}; 
others claim that 85\% of ML projects fail to deliver value~\cite{ai-investments}.

At the same time, it is unclear {\em why} MLOps is hard.
Our present-day understanding of MLOps is limited to a fragmented landscape of white papers, anecdotes, and 
thought pieces~\cite{eric, mlreef_2021, garcia2018context, ghosh_2021, mlopsoda, gartnerinc},
as well as a cottage industry of startups aiming to address MLOps issues~\cite{huyen_2020}.
Early work by Sculley et al.
attributes MLOps challenges to ``technical debt'',
due to which there is ``massive ongoing maintenance costs
in real-world ML systems''~\cite{Sculley2015HiddenTD}.
Most successful ML deployments seem to involve
a ``team of engineers who spend a
significant portion of their time on the less 
glamorous aspects of ML like maintaining and monitoring
ML pipelines''~\cite{polyzotis2017data}.
Prior work has studied general practices of 
data analysis and science~\cite{sambasivan2021everyone, kandel, muller2019data, zhang2020data}, 
without considering MLOps challenges of productionizing models. 

There is thus a pressing need to bring clarity to MLOps, specifically 
in identifying what MLOps typically involves---across organizations and ML applications.
A richer understanding of best practices and challenges in MLOps can surface gaps 
in present-day processes and better inform the development of next-generation tools. Therefore, we conducted a semi-structured interview study of ML engineers (MLEs), each of whom has worked on ML models in production. 
We sourced 18 participants from different organizations and applications (\Cref{tab:interviewees}) and asked them open-ended questions to understand their workflow and day-to-day challenges.

\begin{figure}
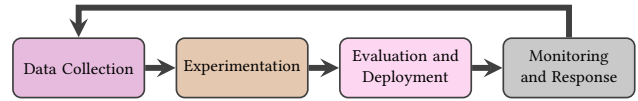

    \centering
         \tikzset{
  every shadow/.style={
    fill=none,
    shadow xshift=0pt,
    shadow yshift=0pt}
}
\tikzset{module/.append style={top color=\col,bottom color=\col}}
    \smartdiagramset{
    set color list={datacoll!50, expt!50, evaldeploy!50, monitorresp!50},
    uniform arrow color=true,
    arrow color=gray!50!black,
    border color=gray!50!black,
    text width=2cm
    }
    \resizebox{\linewidth}{!}{
    \smartdiagram[flow diagram:horizontal]{Data Collection, Experimentation, Evaluation and Deployment, Monitoring and Response}}
    \vspace{-15pt}
    \caption{Routine tasks in the ML engineering workflow.}
    \label{fig:tasksdiagram}
    \vspace{-15pt}
\end{figure}

We find that MLEs perform {\bf four routine tasks}, shown in \Cref{fig:tasksdiagram}: (i) data collection, (ii) experimentation, (iii) evaluation and deployment, and (iv) monitoring and response. Across tasks, we observe {\bf three variables} that dictate success for a production ML deployment: Velocity, Validation, and Versioning.\footnote{Our Three Vs of MLOps aren't meant to be confused with the Three Vs of Big Data (Volume, Variety, Velocity)~\cite{russom2011big}. The first authors learned of the Big Data Vs after drafting the MLOps Vs and were surprised to find similarities around volume/versioning and velocity.} We describe common MLOps practices, grouped under overarching findings: 

\topic{ML engineering is very experimental in nature (\Cref{sec:summary-expt})} As mentioned earlier, various articles claim that it is a problem for 90\% of models to never make it to production~\cite{weiner2020ai}, but we find that this statistic is misguided. 
The nature of constant experimentation is bound to create many versions, 
a small fraction of which (i.e. ``the best of the best'') 
will make it to production. 
Thus it is beneficial to prototype ideas quickly, by making minimal changes to existing workflows, and demonstrate practical benefits early---so that bad models never make it far. 

\topic{Operationalizing model evaluation requires an active organizational effort (\Cref{sec:summary-eval})} Popular model evaluation ``best practices'' do not do justice to the rigor with which organizations think about deployments: they generally focus on using one typically-static held-out dataset to evaluate the model on~\cite{lin2021the} and a single ML metric choice (e.g., precision, recall)~\cite{courseraeval, courserasingnumb}. We find that MLEs invest significant resources in maintaining multiple up-to-date evaluation datasets and metrics over time---especially ensuring that data sub-populations of interest are adequately covered.

\topic{Non-ML rules and human-in-the-loop practices keep models reliable in production (\Cref{sec:summary-hacks})} We find that MLEs prefer simple ideas, even if it means handling multiple versions: for example, rather than leverage advanced techniques to minimize distribution shift errors~\cite{rethinkingiw, arm}, MLEs would simply create new models, retrained on fresh data. MLEs ensured that deployments were reliable via strategies such as on-call rotations, model rollbacks, or elaborate rule-based guardrails to avoid incorrect outputs.

In \Cref{sec:challenges}, we discuss recurring MLOps challenges across all tasks. We express these pain points as tensions and synergies between our three ``V'' variables---for example, undocumented ``tribal knowledge'' about pipelines (\Cref{sec:challenges-antipatterns-tribal}) demonstrates a tension between velocity (i.e., quickly changing the pipeline in response to a bug) and well-executed versioning (i.e., documenting every change). We conclude the description of each pain point with a discussion of opportunities for future tools.

\begin{table}[]
    \centering
    \begin{tabular}{r|c|c|c}
    \toprule
        & \textbf{Role}   & \textbf{Company Size} & \textbf{Application}    \\
       \midrule
p1           & MLE Manager     & Large                 & Autonomous vehicles       \\
p2           & MLE             & Medium                & Autonomous vehicles       \\
p3           & MLE & Small                 & Computer hardware           \\
p4           & MLE             & Medium                & Retail                  \\
p5           & MLE Manager     & Large                 & Ads                     \\
p6           & MLE             & Large                 & Cloud computing         \\
p7           & MLE             & Small                 & Finance                 \\
p8           & MLE             & Small                 & NLP                     \\
p10          & MLE             & Small                 & OCR + NLP               \\
p11          & MLE Manager     & Medium                & Banking                 \\
p12          & MLE             & Large                 & Cloud computing         \\
p13          & MLE              & Small                 & Bioinformatics          \\
p14          & MLE             & Medium                & Cybersecurity      \\
p15          & MLE             & Medium                & Fintech                 \\
p16          & MLE              & Small                 & Marketing and analytics \\
p17          & MLE              & Medium                & Website builder         \\
p18          & MLE             & Large                 & Recommender systems     \\
p19          & MLE Manager     & Large                 & Ads        \\            \bottomrule
    \end{tabular}
    \caption{Anonymized description of interviewees. Small companies have fewer than 100 employees; medium-sized companies have 100-1000 employees, and large companies have 1000 or more employees.}
    \label{tab:interviewees}
\end{table}

\section{Related Work}
\label{sec:related}

Several books and papers in the traditional software engineering literature describe the need for DevOps, a combination of software \emph{dev}elopers and \emph{op}erations teams, to streamline the process of delivering software in organizations~\cite{leite2019survey, ebert2016devops, lwakatare2019devops, loukides2012devops}. Similarly, \emph{MLOps}, or DevOps principles applied to machine learning, has emerged from the rise of machine learning (ML) application development in software organizations. MLOps is a nascent field, where most existing papers give definitions and overviews of MLOps, as well as its relation to ML, software engineering, DevOps, and data engineering~\cite{mlopsoda, makinen2021needs, tamburri2020sustainable, john2021towards}. MLOps poses unique challenges because of its focus on developing, deploying, and sustaining models, or artifacts that need to reflect data as data changes over time~\cite{renggli2021data, Shankar2022TowardsOF, Shankar2022RethinkingSM}. We discuss work related to MLOps workflows, challenges and interview studies for ML.

\topic{MLOps Workflow} The MLOps workflow involves supporting data collection and processing, experimentation, evaluation and deployment, and monitoring and response, as shown in~\Cref{fig:tasksdiagram}. Several research papers and companies have proposed tools to accomplish various tasks in the workflow, such as data pre-processing~\cite{ratner2017snorkel, mlinspect, dagger} and experiment tracking~\cite{mlflow, wandb, modeldb}. 
Crankshaw et al. studied the problem of model deployment and low-latency prediction serving~\cite{crankshaw2017clipper}. 
With regards to validating changes in \emph{production} systems, some researchers have studied CI (Continuous Integration) for ML and proposed preliminary solutions---for example, \texttt{ease.ml/ci} streamlines data management and proposes unit tests for overfitting~\cite{aguilar2021ease}, Garg et al. survey different MLOps tools~\cite{mlopssurvey}, and some papers introduce tools to perform validation and monitoring in production ML pipelines~\cite{dataval, schelter, kang2018model}. 

\topic{MLOps Challenges} 
Sculley et al. were early proponents that production ML systems raise special challenges and can be hard to maintain over time, based on their experience at Google~\cite{Sculley2015HiddenTD}. Since then, several research projects have emerged to explore and tackle individual challenges in the MLOps workflow. 
For example, some discuss the need to manage data provenance and training context for model debugging  purposes~\cite{namaki2020vamsa, vizier, hellerstein2017ground, garcia2020flor}. Others describe the challenges of handling state
and ensuring reproducibility (i.e., ``managing messes'') while using computational notebooks~\cite{nbsafety, nbslicer, nbgather}. Additionally, data distribution shifts have been technically but not operationally studied---i.e., how humans debug such shifts in practice~\cite{failingloudly, Ovadia2019CanYT, unifyingview, Wiles2021AFA, sugiyama}. Rather than focus on a single pain point, Lee et al. analyze challenges across ML workflows on an open-source ML platform~\cite{demystifying}. Similarly, Xin et al.~\cite{xin2021production}
analyze ML pipelines at Google to understand typical model configurations and retraining patterns. Polyzotis et al.~\cite{polyzotis_data_2018,polyzotis2017data} survey
challenges centric to data management for machine learning deployments.
Paleyes et al. review published reports of individual ML deployments and survey common challenges~\cite{deploychallenges}. 
Our study instead focuses on issues across the production workflow (i.e., MLOps practices and challenges) as opposed to individual pain-points,
identified by interviewing 
those who are are most affected by it---the ML engineers.

\topic{Data Science and ML-Related Interview Studies} Kandel et al.~\cite{kandel} interview data analysts at enterprises, focusing on broader organizational contexts like we do; however, MLOps workflows and challenges extend beyond data analysis. 
Other studies build on Kandel et al.'s work, exploring aspects such as collaboration,
code practices, and tools~\cite{muller2019data, zhang2020data, kery2017variolite, kim2017data, passi2018trust}, all centered on general data analysis and data science,
as opposed to transitioning workflows in ML to production.
Many ML-related interview studies focus on a single tool, task, or challenge in the workflow---for example, AutoML~\cite{autoai, whitherautoml}, data iteration~\cite{hohman2020understanding}, model training~\cite{ensemblematrix}, minimizing bias in ML models~\cite{law2020designing, holsteinfairness, Madaiochecklists}, and building infrastructure for ML pipelines~\cite{muiruri2022practices}. 
Sambasivan et al.~\cite{sambasivan2021everyone} study data quality issues during machine learning, as opposed to challenges in MLOps. 
Other ML-related interview studies focus on specific applications of ML, such as medicine~\cite{pumplun2021adoption}, customer service~\cite{folstad2018makes}, and interview processing~\cite{billington2022machine}. Some interview studies report on software engineering practices for ML development; however, they focus only on a few applications and primarily on engineering, not operational, challenges~\cite{msoftcasestudy, Lwakatare2019ATO}. Our interview study aims to be both broad and focused: we consider many applications and companies, but is centered around the engineers that perform MLOps tasks, with an eye towards highlighting both engineering and operational practices and challenges. Additionally, our focus is  on learning how models are deployed and sustained in production---we discover this by interviewing ML practitioners directly.

\section{Methods}
\label{sec:methods}

Following review by our institution's review board,
we conducted an interview study of 18 ML Engineers (MLEs) 
working across a wide variety of sectors to learn 
more about their first-hand experiences 
serving and maintaining models in production.

\subsection{Participant Recruitment}\label{participant_recruitment}
We recruited persons
who were responsible for the development, regular retraining, monitoring and deployment of any ML model \emph{in production}. 
A description of the 18 MLEs (22\% female-identifying\footnote{The skewed gender distribution is one possible indicator of sampling bias. We openly and actively recruited female-idenitifying MLEs to mitigate some sampling bias.}) is shown in~\Cref{tab:interviewees}.
The MLEs we interviewed varied in their educational backgrounds, years of experience, roles, team size, and work sector. 
Recruitment was conducted in rounds 
over the course of an academic year (2021-2022).
In each round, between three to five candidates were reached by email and invited to participate. 
We relied on our professional networks and open calls posted on MLOps channels in Discord\footnote{mlops.discord.com}, Slack\footnote{mlops-community.slack.com}, and Twitter to compile a roster of candidates.
The roster was incrementally updated roughly after every round of interviews, integrating information gained from the concurrent coding and analysis of transcripts (Section~\ref{coding_and_analysis}).
Recruitment rounds were repeated until 
we reached saturation on our findings~\cite{muller2014curiosity}.

\subsection{Interview Protocol}\label{interview_protocol}

With each participant, we 
conducted semi-structured interviews 
over video call lasting 45 to 75 minutes each.
Over the course of the interview, we asked descriptive, structural, and contrast questions abiding by ethnographic interview guidelines~\cite{spradley2016ethnographic}. The questions are listed in \Cref{app:interviewq}.
Specifically, our questions spanned six categories: 
(1) the type of ML task(s) they work on; 
(2) the approach(es) they use for developing and experimenting on models;
(3) how and when they transition from development/experimentation
to deployment;
(4) how they evaluate their models prior to deployment;
(5) how they monitor their deployed models; and
(6) how they respond to issues or bugs 
that may emerge during deployment
or otherwise. We covered these categories to get an overarching understanding of an ML lifecycle deployment, from conception to production sustenance.

Participants received a written consent form before the interview, and agreed to participate free of compensation.
As per our agreement, we automatically transcribed the interviews using Zoom software. 
In the interest of privacy and confidentiality, we did not record audio or video of the interviews. Transcripts were redacted of personally identifiable information before being uploaded to a secured drive in the cloud. More information about the transcripts can be found in \Cref{app:transcripts}.

\subsection{Transcript Coding \& Analysis}\label{coding_and_analysis}

Following a grounded theory approach~\cite{strauss1994grounded,cho2014reducing},
we employed open and axial coding to analyze our transcripts. 
We used MaxQDA, a common qualitative analysis software package for coding and comparative analysis.
During a coding pass, two study personnel
independently read interview transcripts closely
to group passages into codes or categories. 
Coding passes were either top-down or bottom-up, meaning that codes were derived from theory or induced from interview passages, respectively. 
Between coding passes, 
study personnel met to discuss surprises and other findings, and following consensus, the code system was revised to reflect changes to the emerging theory. 
Coding passes were repeated until reaching convergence. More information about the codes is shown in \Cref{app:codes}, including a list of the most frequently occurring codes (\Cref{tab:freqcodes}) and co-occurring codes (\Cref{fig:correlatedcodes}).

\section{MLOps Practices: Our Findings}
\label{sec:summary}

In this section, we present information about common practices in production ML deployments that we learned from the interviews. First, we describe common tasks in the production ML workflow in \Cref{sec:summary-tasks}. Next, we introduce the Three Vs of MLOps, grounding both the discussion of findings and the challenges that we will explain in ~\Cref{sec:challenges}. Then in ~\Cref{sec:summary-expt}, we describe the strategies ML engineers leverage to produce successful experiment ideas. In ~\Cref{sec:summary-eval}, we discuss organizational efforts to effectively evaluate models. Finally, in ~\Cref{sec:summary-hacks}, we investigate the hacks ML engineers use to sustain high performance in productions ML pipelines. 

\subsection{Tasks in the Production ML Lifecycle}
\label{sec:summary-tasks}

We characterized ML engineers' workflows into four high-level tasks, each of which employ a wide variety of tools. We briefly describe each task in turn, and elaborate on them
as they arise in our findings below.

\topic{Data Collection and Labeling} Data collection spans sourcing new data, wrangling data from sources into a centralized repository, and cleaning data. Data labeling can be outsourced (e.g., Mechanical Turk) or performed in-house with teams of annotators. Since descriptions and interview studies of data collection, analysis, wrangling and labeling activities can be found in related papers~\cite{kandel2011wrangler, kandel, datacleaning, sambasivan2021everyone}, we focus our summary of findings on the other three tasks. 

\topic{Feature Engineering and Model Experimentation} ML engineers typically focus on improving ML performance, measured via metrics such as accuracy or mean-squared-error. Experiments can be data-driven or model-driven; for example, an engineer can create a new feature or change the model architecture from tree-based to neural network-based. 

\topic{Model Evaluation and Deployment} A model is typically evaluated by computing a metric (e.g., accuracy) over a collection of labeled data points hidden at training time, or a \emph{validation dataset}, to see if its performance is better than what the currently-running production model achieved during its evaluation phase. Deployment involves reviewing the proposed change, possibly staging the change to increasing percentages of the population, or A/B testing on live users, and keeping records of any change to production in case of a necessary rollback. 

\topic{ML Pipeline Monitoring and Response} Monitoring ML pipelines and responding to bugs involve tracking live metrics (via queries or dashboards), slicing and dicing sub-populations to investigate prediction quality, patching the model with non-ML heuristics for known failure modes, and finding in-the-wild failures and adding them to the evaluation set. 

\subsection{Three Vs of MLOps: Velocity, Validation, Versioning}
\label{sec:3vs}


When developing and pushing ML models to production, three properties of the workflow and infrastructure dictate how successful deployments will be: Velocity, Validation, and Versioning, discussed in turn.

\topic{Velocity} Since ML is so experimental in nature, it's important to be able to prototype and iterate on ideas quickly (e.g., go from a new idea to a trained model in a day). ML engineers attributed their productivity to development environments that prioritized high experimentation velocity and debugging environments that allowed them to test hypotheses quickly (P1, P3, P6, P10, P11, P14, P18). 

\topic{Validation} Since errors become more expensive to handle when users see them, it's good to test changes, prune bad ideas, and proactively monitor pipelines for bugs as early as possible (P1, P2, P5, P6, P7, P10, P14, P15, P18). P1 said: \blockquote{The general theme, as we moved up in maturity, is: how do you do more of the validation earlier, so the iteration cycle is faster?}

\topic{Versioning} Since it's impossible to anticipate all bugs before they occur, it's helpful to store and manage multiple versions of production models and datasets for querying, debugging, and minimizing production pipeline downtime. ML engineers responded to buggy models in production by switching the model to a simpler, historical, or retrained version (P6, P8, P10, P14, P15, 18).


\subsection{Machine Learning Engineering is Very Experimental, Even in Production}
\label{sec:summary-expt}

ML engineering, as a discipline, is highly experimental and iterative in nature, especially compared to typical software engineering. Contrary to popular negative sentiment around the large numbers of experiments and models that don't make it to production, we found that it's actually okay for experiments and models not to make it to production. What matters is making sure ideas can be prototyped and validated quickly---so that bad ones can be pruned away immediately. While there is no substitute for on-the-job experience to learn how to choose successful projects (P5), we document some self-reported strategies from our interviewees. 

\subsubsection{Good project ideas start with collaborators.} Project ideas, such as new features, came from or were validated early by domain experts, data scientists and analysts who had already performed a lot of exploratory data analysis. P14 and P17 independently recounted successful project ideas that came from asynchronous conversations on Slack: P17 said, \blockquote{I look for features from data scientists, [who have ideas of] things that are correlated with what I'm trying to predict.} Solely relying on other collaborators wasn't enough, though---P5 mentioned that they \blockquote{still need to be pretty proactive about what to search for.}

Some organizations explicitly prioritized cross-team collaboration as part of their culture. P11 said:

\begin{displayquote}
We really think it's important to bridge that gap between what's often, you know, a [subject matter expert] in one room annotating and then handing things over the
wire to a data scientist---a scene where you have no communication. So we make sure there's both data science and subject matter expertise representation [on our teams].
\end{displayquote}

To foster a more collaborative culture, P16 discussed the concept of ``building goodwill'' with other teams through tedious tasks that weren't always explicitly a part of company plans: \blockquote{Sometimes we'll fix something [here and] there to like build some good goodwill, so that we can call on them in the future...I do this stuff as I have to do it, not because I'm really passionate about doing it.} 

\subsubsection{Iterate on the data, not necessarily the model.} Several participants recommended focusing on experiments that provide additional context to the model, typically via new features (P5, P6, P11, P12, P14, P16, P17, P18, P19).  P17 mentioned that most ML projects at their organization centered around adding new features. P14 mentioned that one of their current projects was to move feature engineering pipelines from Scala to SparkSQL (a language more familiar to ML engineers and data scientists), so experiment ideas could be coded and validated faster. P11 noted that iterating on the data, not the model, was preferable because it resulted in faster velocity:

\begin{displayquote}
I'm gonna start with a [fixed] model because it means faster [iterations]. And often, like most of the time empirically, it's gonna be something in our data that we can use to kind of push the boundary...obviously it's not like a dogmatic We Will Never Touch The Model, but it shouldn't be our first move.
\end{displayquote}
Prior work has also identified the importance of data work~\cite{sambasivan2021everyone}.

\subsubsection{Account for diminishing returns.}\label{sec:summary-expt-diminishing} At many organizations (especially larger companies), deployment can occur in stages---i.e., first validated offline, then deployed to 1\% of production traffic, then validated again before a deployment to larger percentages of traffic. Some interviewees (P5, P6, P18) explained that experiment ideas typically have diminishing performance gains in later stages of deployment. As a result, P18 mentioned that they would initially try multiple ideas but focus only on ideas with the largest performance gains in the earliest stage of deployment; they emphasized the importance of \blockquote{validat[ing] ideas offline...[to make] productivity higher.} P19 corroborated this by saying end-to-end staged deployments could take several months, making it a high priority to kill ideas with minimal gain in early stages to avoid wasting future time. Additionally, to help with validating early, many engineers discussed the importance of a sandbox for stress-testing their ideas (P1, P5, P6, P11, P12, P13, P14, P15, P17, P18, P19). For some engineers, this was a single Jupyter notebook; others' organizations had separate sandbox environments to load production models and run ad-hoc queries. 

\subsubsection{Small changes are preferable to larger changes.} In line with software best practices, interviewees discussed keeping their code changes as small as possible for multiple reasons, including faster code review, easier validation, and fewer merge conflicts (P2, P5, P6, P10, P11, P18, P19). Additionally, changes in large organizations were primarily made in config files instead of main application code (P1, P2, P5, P10, P12, P19). For example, instead of editing parameters directly in a Python model training script, it was preferable to edit a config file (e.g., JSON or YAML) of parameters instead and link the config file to the model training script. 

P19 described how, as their team matured, they edited the model code less: \blockquote{Eventually it was the [DAG] pipeline code which changed more...there was no reason to touch the [model] code...everything is config-based.} P5 mentioned that several of their experiments involved \blockquote{[taking] an existing model, modify[ing] [the config] with some changes, and deploying it within an existing training cluster.} Supporting a config-driven development was important, P1 said, otherwise bugs might arise when promoting the experiment idea to production:

\begin{displayquote}
People might forget to, when they spawn multiple processes, to do data loading in parallel, they might forget to set different random seeds, especially [things] you have to do explicitly many times...you're talking a lot about these small, small things you're not going to be able to catch [at deployment time] and then of course you won't have the expected performance in production.
\end{displayquote}

Because ML experimentation requires many considerations to yield correct results---e.g., setting random seeds, accessing the same versions of code libraries and data---constraining engineers to config-only changes can reduce the number of bugs.

\subsection{Operationalizing Model Evaluation is an Active Effort}
\label{sec:summary-eval}

We found that MLEs described intensive model evaluation efforts at their companies to keep up with data changes, product and business requirement changes, user changes, and organizational changes. The goal of model evaluation is to prevent repeated failures and bad models from making it to production while maintaining velocity---i.e., the ability for pipelines to quickly adapt to change.

\subsubsection{Validation datasets should be dynamic.}

Many engineers reported processes to analyze live failure modes and update the validation datasets to prevent similar failures from happening again (P1, P2, P5, P6, P8, P11, P15, P16, P17, P18). P1 described this process as a departure from what they had learned in academia: \blockquote{You have this classic issue where most researchers are evaluat[ing] against fixed data sets...[but] most industry methods change their datasets.} We found that these dynamic validation sets served two purposes: (1) the obvious goal of making sure the validation set reflects live data as much as possible, given new learnings about the problem and shifts in the aggregate data distribution, and (2) the more subtle goal of addressing localized shifts that subpopulations may experience (e.g., low accuracy for a specific label). 

The challenge with (2) is that many subpopulations are typically unforeseen; many times they are discovered post-deployment. To enumerate them, P11 discussed how they systematically bucketed their data points based on the model's error and created validation sets for each underperforming bucket:

\begin{displayquote}
Some [of the metrics in my tool] are standard, like a confusion matrix, but it's not really effective because it doesn't drill things down [into specific subpopulations that users care about]. Slices are user-defined, but sometimes it's a little bit more automated. [During offline evaluation, we] find the error bucket that [we] want to drill down, and then [we] either improve the model in very systematic ways or improve [our] data in very systematic ways.
\end{displayquote}


Rather than follow an anticipatory approach of constructing different failure modes in the offline validation phase---e.g., performance drops in subpopulations users might care deeply about---like P11 did, P8 offered a reactive strategy of spawning a new dataset for each observed live failure: \blockquote{Every [failed prediction] gets into the same queue, and 3 of us sit down once a week and go through the queue...then our [analysts] collect more [similar] data.} This new dataset was then used in the offline validation phase in future iterations of the production ML lifecycle.

While processes to dynamically update the validation datasets ranged from human-in-the-loop to frequent synthetic data construction (P6), we found that higher-stakes applications of ML (e.g., autonomous vehicles), created separate teams to manage the dynamic evaluation process. P1 said:

\begin{displayquote}
 We had to move away from only aggregate metrics like MAP towards the ability to curate scenarios of interest, and then validate model performance on them specifically.
So, as an example, you can't hit pedestrians, right. You can't hit cyclists. You need to work in roundabouts. You have a base layer of ML performance and the performance is not perfect...what you need to be able to do in a mature MLOps pipeline is go very quickly from user recorded bug, to not only are you going to fix it, but you also have to be able to drive improvements to the stack by changing your data based on those bugs.
\end{displayquote}

Although the dynamic evaluation process might require many humans in the loop---a seemingly intense organizational effort---engineers thought it was crucial to have. When asked why they invested a lot of energy into their dynamic process, P11 said: \blockquote{I guess it was always a design principle---the data is [always] changing.}

\subsubsection{Validation systems should be standardized.}

The dynamic nature of validation processes makes it hard to effectively maintain versions of such processes, motivating efforts to standardize them. Several participants recalled instances of bugs stemming from inconsistent definitions of successful validation---i.e., where different engineers on their team evaluated models differently, causing unexpected changes to live performance metrics (P1, P3, P4, P5, P6, P7, P17). For instance, P4 lamented that every engineer working on a particular model had a cloned version of the main evaluation notebook, with a few changes. The inconsistent requirements for promoting a model to production caused headaches while monitoring and debugging, so their company instated a new effort to standardize evaluation scripts.

Although other MLOps practices highlighted the synergy between velocity and validating early (\Cref{sec:summary-expt-diminishing}), we found that standardizing the validation system exposed a tension between velocity (i.e., being able to promote models quickly) and validating early, or eliminating the possibility of some bugs at deployment time. Since many validation systems needed to frequently change, as previously discussed, turnaround times for code review and merges to the main branch often could not keep up with the new tests and collections of data points added to the validation system. So, it was easier for engineers to fork and modify the evaluation system. However, P2 discussed that the decrease in velocity was worth it for their organization when they standardized evaluation:

\begin{displayquote}
We have guidelines on how to run eval[uation] comprehensively when any particular change is being made. Now there is a merge queue, and we have to make sure that we process the merge queue in order, and that improvements are actually also reflected in subsequent models, so it requires some coordination. We'd much rather gate deploying a model than deploy a model that's bad. So we tend to be pretty conservative [now].
\end{displayquote}

A standardized evaluation system also reduced friction in deploying ML in large companies and high-stakes settings. P5 discussed that for some models, deployments needed approvals across the organization, and it was much harder to justify a deployment with a custom or ad-hoc evaluation process: \blockquote{At the end of the day, it's all a business-driven decision...for something that has so much revenue riding on it, [you can't have] a subjective opinion on whether [your] model is better.}

\subsubsection{Spread a deployment across multiple stages, and evaluate at each stage.}

Several organizations, particularly those with many customers, had a multi-stage deployment process for new models or model changes, progressively evaluating at each stage (P3, P5, P6, P7, P8, P12, P15, P17, P18, P19). P6 described the staged deployment process as:

\begin{displayquote}
In [the large companies I've worked at], when we deploy code it goes through what's
called a staged deployment process, where we have designated test clusters, [stage 1] clusters, [stage 2] clusters, then the global deployment [to all users]. The idea here is you deploy increasingly along these clusters, so that you catch problems before they've met customers.
\end{displayquote}

Each organization had different names for its stages (e.g., test, dev, canary, staging, shadow, A/B) and different numbers of stages in the deployment process (usually between one and four). The stages helped invalidate models that might perform poorly in full production, especially for brand-new or business-critical pipelines. P15 recounted an initial chatbot product launch using their staged deployment process, claiming it successfully "made it" because they were able to catch failures and update the model in early stages:

\begin{displayquote}
We spent a long time very slowly, ramping up the model to very small percentages of traffic and watching what happened. [When there was a failure mode,] a product person would ping us and say: hey, this was kind of weird, should we create a rule around this [suggested text] to filter this out?
\end{displayquote}


Of particular note was one type of stage, the shadow stage---where predictions were generated live but not surfaced to users, that came before a deployment to a small fraction of live users. P14 described how they used the shadow stage to assess how impactful new features could be:

\begin{displayquote}
So if we're testing out a new idea and want to see, what would the impact be for this new set of features without actually deploying that into production, we can deploy that in a type of shadow mode where it's running alongside the production model and making predictions. We track all the metrics for [both] models in [our data lake]...so we can compare them easily.
\end{displayquote}

Shadow mode had other use cases---for instance, P15 discussed how shadow mode was used to convince other stakeholders (e.g., product managers, business analysts) that a new model or change to an existing model was worth putting in production. P19 mentioned that they use shadow mode to invalidate experiment ideas that would eventually fail. But shadow mode alone wasn't a substitute for all stages of deployment---P6 said, \blockquote{[in the early stage], we don't have a good sample of how the model is going to behave in production}---requiring the multiple stages. Additionally, in products that have a feedback loop (e.g., recommender systems), it is impossible to evaluate the model in shadow mode because users do not interact with shadow predictions.

\subsubsection{ML evaluation metrics should be tied to product metrics.}

Multiple participants stressed the importance of evaluating metrics critical to the product, such as click-through rate or user churn rate, rather than ML-specific metrics alone like MAP (P5, P7, P15, P16, P11, P17, P18, P19). The need to evaluate product-critical metrics stemmed from close collaboration with other stakeholders, such as product managers and business operators. P11 felt that a key reason many ML projects fail is that they don't measure metrics that will yield the organization value:

\begin{displayquote}
Tying [model performance] to the business's KPIs (key performance indicators) is really important. But it's a process---you need to figure out what [the KPIs] are, and frankly I think that's how people should be doing AI. It [shouldn't be] like: hey, let's do these experiments and get cool numbers and show off these nice precision-recall curves to our bosses and call it a day. It should be like: hey, let's actually show the same business metrics that everyone else is held accountable to to our bosses at the end of the day.
\end{displayquote}

Since product-specific metrics are, by definition, different for different ML models, it was important for engineers to treat choosing the metrics as an explicit step in their workflow and align with other stakeholders to make sure the right metrics were chosen. For example, P16 said that for every new ML project they work on, their \blockquote{first task is to figure out, what are customers actually interested in, or what's the metric that they care about.} P17 said that every model change in production is validated by the product team: \blockquote{if we can get a statistically significant greater percentage [of] people to subscribe to [the product], then [we can fully deploy].}

For some organizations, a consequence of tightly coupling evaluation to product metrics was an additional emphasis on important customers during evaluation (P6, P10). P6 described how, at their company, experimental changes that increased aggregate metrics could sometimes be prevented from going to production:

\begin{displayquote}
There's an [ML] system to allocate resources for [our product]. We have hard-coded rules for mission critical customers. Like at the beginning of Covid, there were hospital [customers] that we had to save [resources] for.
\end{displayquote}

Participants who came from research or academia noted that tying evaluation to the product metrics was a different experience. P6 commented:

\begin{displayquote}
I think about where the business will benefit from what we're building. We're not just shipping off fake wins, like we're really in the value business. You've got to see value from AI in your organization in order to feel like [our product] was worth it to you, and I guess that's a mindset that we really ought to have [as a broader community].
\end{displayquote}

\subsection{Sustaining Models Requires Deliberate Software Engineering and Organizational Practices}
\label{sec:summary-hacks}

Here, we present a list of strategies ML engineers employed during monitoring and debugging phases to sustain model performance post-deployment. 

\subsubsection{Create new versions: frequently retrain on and label live data.} Production ML bugs can be detected by tracking pipeline performance, measured by metrics like prediction accuracy, and triggering an alert if there is a drop in performance that exceeds some predefined threshold. On-call ML engineers noted that reacting to an ML-related bug in production often took a long time, motivating them to find alternative strategies to quickly restore performance (P1, P7, P8, P10, P14, P15, P17, P19). P14 mentioned automatically retraining the model every day so model performance would not suffer for longer than a day:

\begin{displayquote}
Why did we start training daily? As far as I'm aware, we wanted to start simple---we could just have a single batch job that processes new data and we wouldn't need to worry about separate retraining schedules. You don't really need to worry about if your model has gone stale if you're retraining it every day.
\end{displayquote}

Retraining cadences ranged from hourly (P18) to every few months (P17) and were different for different models within the same organization (P1). None of the participants interviewed reported any scientific procedure for determining the cadence; the retraining cadences were set in a way that streamlined operations for the organization in the easiest way. For example, P18 mentioned that \blockquote{retraining takes about 3 to 4 hours, so [they] matched the cadence with it such that as soon as [they] finished any one model, they kicked off the next training [job].}

Some engineers reported an inability to retrain unless they had freshly labeled data, motivating their organizations to set up a team to frequently label live data (P1, P8, P10, P11, P16). P10 reported that a group within their company periodically collected new documents for their language models to fine-tune on. P11 mentioned an in-house team of junior analysts to annotate the data; however, a problem was that these annotations frequently conflicted and the organization did not know how to reconcile the noise.

\subsubsection{Maintain old versions as fallback models.} Another way to minimize downtime when a model is known to be broken is to have a fallback model to revert to---either an old version or simple version. P19 said: \blockquote{if the production model drops and the calibration model is still performing within a [specified] range, we'll fall back to the calibration model until someone will fix the production model.} P18 mentioned that it was important to keep some model up and running, even if they \blockquote{switched to a less economic model and had to just cut the losses.} 

\subsubsection{Maintain layers of heuristics.} P14 and P15 each discussed how their models are augmented with a final, rule-based layer to keep live predictions more stable. For example, P14 mentioned working on an anomaly detection model and adding a heuristics layer on top to filter the set of anomalies that surface based on domain knowledge. P15 discussed one of their language models for a customer support chatbot:

\begin{displayquote}
The model might not have enough confidence in the suggested reply, so we don't return [the recommendation]. Also, language models can say all sorts of things you don't necessarily want it to---another reason that we don't show suggestions. For example, if somebody asks when the business is open, the model might try to quote a time when it thinks the business is open. [It might say] ``9 am'', but the model doesn't know that. So if we detect time, then we filter that [reply]. We have a lot of filters. 
\end{displayquote}

Constructing such filters was an iterative process---P15 mentioned constantly stress-testing the model in a sandbox, as well as observing suggested replies in early stages of deployment, to come up with filter ideas. Creating filters was a more effective strategy than trying to retrain the model to say the right thing; the goal was to keep some version of a model working in production with little downtime. This combination of modern model-driven ML and old-fashioned rule-based AI indicates a need for managing filters (and versions of filters) in addition to managing learned models. The engineers we interviewed managed these artifacts themselves.

\subsubsection{Validate data going in and out of pipelines.}\label{sec:summary-hacks-dataval} While participants reported that model parameters were typically "statically" validated before deploying to full production, features and predictions were continuously monitored for production models (P1, P2, P6, P8, P14, P16, P17, P18, P19). Several metrics were monitored---P2 discussed hard constraints for feature columns (e.g., bounds on values), P6 talked about monitoring completeness (i.e., fraction of non-null values) for features, P16 mentioned embedding their pipelines with "common sense checks," implemented as hard constraints on columns, and P8 described schema checks---making sure each data item adheres to an expected set of columns and their types.

While rudimentary data checks were embedded in most systems, P6 discussed that it was hard to figure out what higher-order data checks to compute:

\begin{displayquote}
Monitoring is both metrics and then a predicate over those metrics that triggers alerts. That second piece doesn't exist---not because the infrastructure is hard, but because no one knows how to set those predicate values...for a lot of this stuff now, there's engineering headcount to support a team doing this stuff. This is people's jobs now; this constant, periodic evaluation of models.
\end{displayquote}

Some participants discussed using black-box data monitoring services but lamented that their alerts did not prevent failures (P7, P14). P7 said:

\begin{displayquote}
We don't find those metrics are useful. I guess, what's the point in tracking these? Sometimes it's really to cover my ass. If someone [hypothetically] asked, how come the performance dropped from X to Y, I could go back in the data and say, there's a slight shift in the user behavior that causes this. So I can do an analysis of trying to convince people what happened, but can I prevent [the problem] from happening? Probably not. Is that useful? Probably not.
\end{displayquote}

While basic data validation was definitely useful for the participants, many of the participants expressed pain points with existing techniques and solutions, which we discuss further in~\Cref{sec:challenges-pain-dataval}.

\subsubsection{Keep it Simple.} Many participants expressed an aversion to complexity, preferring to rely on simple models and algorithms whenever possible (P1, P2, P6, P7, P11, P12, P14, P15, P16, P17, P19). P7 described the importance of relying on a simple training and hyperparameter search algorithm:

\begin{displayquote}
In finance, we always split data by time. The thing I [learned in finance] is, don't exactly try to tune the hyperparameters too much, because that just overfits to historic data.
\end{displayquote}

P7 discussed choosing tree-based models over deep learning models for their ease of use, which simplified post-deployment maintenance: \blockquote{I can probably do the same thing with neural nets. But it's not worth it. [After] deployment it just doesn't make any sense at all.} However, other participants chose to use deep learning as a means of simplifying their pipelines (P1, P16). For instance, P16 described training a small number of higher-capacity models rather than a separate model for each target: \blockquote{There were hundreds of products that [customers] were interested in, so we found it easier to instead train 3 separate classifiers that all shared the same underlying embedding...from a neural network.}

While there was no universally agreed-upon answer to a question as broad as, ``should I use deep learning?" we found a common theme in how participants leveraged deep learning models. Specifically, for ease of post-deployment maintenance (e.g., an ability to retroactively debug pipelines), outputs of deep learning models were typically human-interpretable (e.g., image segmentation, object recognition, probabilities or likelihoods as embeddings). P1 described a push at their company to rely \emph{more} on neural networks:

\begin{displayquote}
A general trend is to try to move more into the neural network, and to combine models wherever possible so there are fewer bigger models. Then you don't have these intermediate dependencies that cause drift and performance regressions...you eliminate entire classes of bugs and and issues by consolidating all these different piecemeal stacks.
\end{displayquote}

\subsubsection{Organizationally Supporting ML Engineers Requires Deliberate Practices.} Our interviewees reported various organizational processes for sustaining models as part of their ML infrastructure. P6, P12, P14, P16, P18, and P19 described \emph{on-call processes} for supervising production ML models. For each model, at any point in time, some ML engineer would be on call, or primarily responsible for it. Any bug or incident observed (e.g., user complaint, pipeline failure) would receive a ticket, created by the on-call engineer, and be placed in a queue. On-call rotations lasted a week or two weeks. At the end of a shift, an engineer would create an incident report---possibly one for each bug---detailing major issues that occurred and how they were fixed. 

Additionally, P6, P7, P8, P10, P12, P14, P15, P16, P18, and P19 mentioned having a \emph{central queue of production ML bugs} that every engineer added tickets to and processed tickets from. Often this queue was larger than what engineers could process in a timely manner, so they assigned priorities to tickets. Finally, P6, P7, P10, and P15 discussed having \emph{Service Level Objectives (SLOs)}, or commitments to minimum standards of performance, for pipelines in their organizations. For example, an pipeline to classify images could have an SLO of 95\% accuracy. A benefit of using the SLO framework for ML pipelines is a clear indication of whether a pipeline is performing well or not---if the SLO is not met, the pipeline is broken, by definition.

\section{MLOps Challenges and Opportunities}
\label{sec:challenges}

In this section, we enumerate common pain points and anti-patterns observed across interviews. We discuss each pain point as a tension or synergy between the Three Vs (\Cref{sec:3vs}). At the end of each pain point, we describe our takeaways of ideas for future tools and research. Finally, in \Cref{sec:challenges-caution}, we characterize layers of the MLOps tool stack for those interested in building MLOps tools.

\subsection{Pain Points in Production ML}
\label{sec:challenges-pain}

We focus on four themes that we didn't know before the interviews: the mismatch between development and production environments, handling a spectrum of data errors, the ad-hoc nature of ML bugs, and long validation processes.

\subsubsection{Mismatch Between Development and Production Environments}
\label{sec:challenges-pain-mismatch}

While it is important to create a separate development environment to validate ideas before promoting them to production, it is also necessary to minimize the discrepancies between the two environments. Otherwise, unanticipated bugs might arise in production (P1, P2, P6, P8, P10, P13, P14, P15, P18). {\bf Creating similar development and production environments exposes a tension between velocity and validating early}: development cycles are more experimental and move faster than production cycles; however, if the development environment is significantly different from the production environment, it's hard to validate (ideas) early. 

We discuss three examples of point points caused by the environment mismatch---data leakage, Jupyter notebook philosophies, and code quality.

\topic{Data Leakage} A common issue was \emph{data leakage}---i.e., assuming during training that there is access to data that does not exist at serving time---an error typically discovered after the model was deployed and several incorrect live predictions were made. Anticipating any possible form of data leakage is tedious and hinders velocity; thus, sometimes leakage was retroactively checked during code review (P1). The nature of data leakage ranged across reported bugs---for example, P18 recounted an instance where embedding models were trained on the same split of data as a downstream model, P2 described a \emph{class imbalance} bug where they did not have enough labeled data for a subpopulation at training time (compared to its representation at serving time), and P15 described a bug in which \emph{feedback delay} (time between making a live prediction and getting its ground-truth label) was ignored while training. Different types of data leakage resulted in different magnitudes of ML performance drops: for example, in a pipeline with daily retraining, feedback delays could prevent retraining from succeeding because of a lack of new labels. However, in P18's embedding leakage example, the resulting model was slightly more overfitted than expected, yielding lower-than-expected performance in production but not completely breaking.

\topic{Strong Opinions on Jupyter Notebooks} Participants described strongly opinionated and different philosophies with respect to how to use Jupyter notebooks in their workflows. Jupyter notebooks were heavily used in development to support high velocity, which we did not find surprising. However, we were surprised that although participants generally acknowledged worse code quality in notebooks, some participants preferred to use them in production to minimize the differences between their development and production environments. P6 mentioned that they could debug quickly when locally downloading, executing, and manipulating data from a production notebook run. P18 remarked on the modularization benefits of a migration from a single codebase of scripts to notebooks: 

\begin{displayquote}
We put each component of the pipeline in a notebook, which has made my life so much easier. Now [when debugging], I can run only one specific component if I want, not the entire pipeline... I don't need to focus on all those other components, and this has also helped with iteration.
\end{displayquote}

On the other hand, some participants strongly disliked the idea of notebooks in production (P10, P15). P15 even went as far as to philosophically discourage the use of notebooks in the development environment: \blockquote{Nobody uses notebooks.
Instead, we all work in a shared code base, which is both the training and serving code base and people kick off jobs in the cloud to train models.} Similarly, P10 recounted a shift at their company to move any work they wanted to reproduce or deploy out of notebooks:

\begin{displayquote}
There were all sorts of manual issues. Someone would, you know, run something with the wrong sort of inputs from the notebook, and I'm [debugging] for like a day and a half. Then [I'd] figure out this was all garbage. Eight months ago, we [realized] this was not working. We need[ed] to put in the engineering effort to create [non-notebook] pipelines.
\end{displayquote}

The anecdotes on notebooks identified conflicts between competing priorities: (1) Notebooks support high velocity and therefore need to be in development environments, (2) Similar development and production environments prevents new bugs from being introduced, and (3) It's easy to make mistakes with notebooks in production, e.g., running with the wrong inputs; copy-pasting instead of reusing code. Each organization had different rankings of these priorities, ultimately indicating whether or not they used notebooks in production.

\topic{Non-standardized Code Quality} We found code quality and review practices to be non-standardized and inconsistent across development and production environments. Some participants described organization-wide production coding standards for specific languages (P2, P5), but even the most mature organizations did not have ML-specific coding guidelines for experiments. Generally, experimental code (in development) was not reviewed, but changes to production went through a code review process (P1, P5). Participants felt that code review wasn't too useful, but they did it to adhere to software best practices (P1, P3, P5, P10). P5 mentioned that \blockquote{it's just really not worth the effort; people might catch some minor errors}. P10 hypothesized that the lack of utility came from difficulty of code review:

\begin{displayquote}
It's tricky. You use a little bit of judgment as to where things might go wrong, and you maybe spend more time sort of reviewing that. But bugs will go to production, and [as long as they're not] that catastrophic, [it's okay.]
\end{displayquote}

Code review and other good software engineering practices might make deployments less error-prone. However, because ML is so experimental in nature, they can be significant barriers to velocity; thus, many model developers ignore these practices (P1, P6, P11). P6 said:

\begin{displayquote}
I used to see a lot of people complaining that model developers don't follow software engineering [practices]. At this point, I'm feeling
more convinced that they don't follow software engineering [practices]---[not] because they're lazy, [but because software engineering practices are] contradictory to the agility of analysis and exploration.
\end{displayquote}

\topic{Takeaway} We believe there's an opportunity to create virtualized infrastructure specific to ML needs with similar development and production environments. Each environment should build on the same foundation but supports different modes of iteration (i.e., high velocity in development). Such tooling should also track the discrepancies between environments and minimize the likelihood that discrepancy-related bugs arise.  

\subsubsection{Handling A Spectrum of Data Errors}
\label{sec:challenges-pain-dataval}

As alluded to in~\Cref{sec:summary-hacks-dataval}, we found that ML engineers struggled to handle the spectrum of data errors: hard $\rightarrow$ soft $\rightarrow$ drift (P5, P6, P8, P11, P14, P16, P17, P18, P19). Hard errors are obvious and result in clearly ``bad predictions'', such as when mixing or swapping columns or when violating constraints (e.g., a negative age). Soft errors, such as a few null-valued features in a data point, are less pernicious and can still yield reasonable predictions, making them hard to catch and quantify. Drift errors occur when the live data is from a seemingly different distribution than the training set; these happen relatively slowly over time. One pain point mentioned by the interviewees was that different types of data errors require different responses, and it was not easy to determine the appropriate response. Another issue was that requiring practitioners to manually define constraints on data quality (e.g., lower and upper bounds on values) was not sustainable over time, as employees with this knowledge left the organization.

The most commonly discussed pain point was \emph{false-positive alerts}, or alerts triggered even when the ML performance is adequate. Engineers often monitored and placed alerts on each feature or input column and prediction or output column (P5, P6, P8, P11, P14, P16, P17, P18, P19). Engineers automated schema checks and bounds to catch hard errors, and they tracked distance metrics (e.g., KL divergence) between historical and live features to catch soft and drift errors. If the number of metrics tracked is so large, even with only a handful of columns, the probability that at least one column violates constraints is high! 

Taking a step back, the purpose of assessing data quality \emph{before} serving predictions is to validate early. {\bf Correctly monitoring data quality demonstrates the conflict between validating early and versioning}---if data validation methods flag a broken data point, which in turn rejects the corresponding prediction made by the main ML model, some backup plan or fallback model version (\Cref{sec:summary-hacks}) is necessary. Consequently, an excessive number of false-positive alerts leads to two pain points: (1) unnecessarily maintaining many model versions or simple heuristics, which can be hard to keep track of, and (2) a lower overall accuracy or ML metric, as baseline models might not serve high-quality predictions (P14, P19).

\topic{Dealing with Alert Fatigue} A surplus of false-positive alerts led to fatigue and silencing of alerts, which could miss actual performance drops. P8 said \blockquote{people [were] getting bombed with these alerts.} P14 mentioned a current initiative at their company to reduce the alert fatigue:

\begin{displayquote}
Recently we've noticed that some of these alerts have been rather noisy and not necessarily reflective of events that we care about triaging and fixing.
So we've recently taken a close look at those alerts and are trying to figure out, how can we more precisely specify that query such that it's only highlighting the problematic events?
\end{displayquote}

P18 shared a similar sentiment, that there was \blockquote{nothing critical in most of the alerts.} The only time there was something critical was \blockquote{way back when [they] had to actually wake up in the middle of the night to solve it...the only time [in years].} When we asked what they did about the noncritical alerts and how they acted on the alerts, P18 said:

\begin{displayquote}
You typically ignore most alerts...I guess on record I'd say 90\% of them aren't immediate. You just have to acknowledge them [internally], like just be aware that there is something happening.
\end{displayquote}

The alert fatigue typically materialized when engineers were on-call, or responsible for ML pipelines during a 7 or 14-day shift. P19 recounted how on-call rotations were dreaded amongst their team, particularly for new team members, due to the high rate of false-positive alerts:

\begin{displayquote}
The pain point is dealing with that alert fatigue and the domain expertise necessary to know what to act on during on-call. New members freak out in the first [on-call], so [for every rotation,] we have two members. One member is a shadow, and they ask a lot of questions.
\end{displayquote}

P19 also discussed an initiative at their company to reduce the alert fatigue, ironically with another model:

\begin{displayquote}
The [internal tool] looks at different metrics for what alerts were [acted on] during the on-call...[the internal tool] tries to reduce the noise level, alert. It says, hey, this alert has been populated this like 1,000 times and ignored 45\% of time.
[The on-call member] will acknowledge whether we need to [fix] the issue.
\end{displayquote}

\topic{Creating Meaningful Data Alerts is Challenging} If schema checks and rudimentary column bounds didn't flag all the errors, and distance metrics between historical and live feature values flagged too many false positive errors, how could engineers find a ``Goldilocks'' alert setting?~\footnote{Goldilocks and the Three Bears is a popular Western fairy tale. Goldilocks, the main character, looks for things that are not too big or not too small, things that are ``just right.''} We organized the data-related issues faced by engineers into a hierarchy, from most frequently occurring to least frequently occurring:

\begin{itemize}
    \item \textbf{Feedback delays:} Many participants said that ground-truth labels for live predictions often arrived after a delay, which could vary unpredictably (e.g., human-in-the-loop or networking delays) and thus caused problems for knowing real-time performance or retraining regularly (P2, P7, P8, P15, P17, P18). P7 felt strongly about the negative impact of feedback delays on their ML pipelines:
    
    \begin{displayquote}
    I have no idea how well [models] actually perform on live data. We do log the the [feature and output] data, but feedback is always delayed by at least 2 weeks. Sometimes we might not have feedback...so when we realize maybe something went wrong, it could [have been] 2 weeks ago, and yeah, it's just straight up---we don't even care...nobody is solving the label lag problem. It doesn't make sense to me that a monitoring tool is not addressing this, because [it's] the number one problem.
    \end{displayquote}
    
    P8 discussed how they spent 2-3 \emph{years} developing a human-in-the-loop pipeline to manually label live data as frequently as possible: \blockquote{you want to come up with the rate at which data is changing, and then assign people to manage this rate roughly}. On the other hand, P17 and P19 both talked about how, when they worked on recommender systems, they did not have to experience feedback delay issues. P17 said: \blockquote{With recommendations, it's pretty clear whether or not we got it right because we get pretty immediate feedback. We suggest something, and someone's like go away or they click it.}
    
    \item \textbf{Unnatural data drift:} Often, in production pipelines, data was missing, incomplete, or corrupted, causing model performance to sharply degrade (P3, P6, P7, P10, P16, P17). Several participants cited Covid as an example, but there are other (better) everyday instances of unnatural data drift. P6 described a bug where users had inconsistent definitions of the same word, complicating the deployment of a service to a new user. P7 mentioned a bug where data from users in a certain geographic region arrived more sporadically than usual. P10 discussed a bug where the format of raw data was occasionally corrupted: \blockquote{Tables didn't always have headers in the same place, even though they were the same tables.}
    
    \item \textbf{Natural data drift:} Surprisingly, participants didn't seem too worried about slower, expected natural data drift over time---they noted that frequent model retrains solved this problem (P6, P7, P8, P12, P15, P16, P17). As an anecdote, we asked P17 to give an example of a natural data drift problem their company faced, and they could not think of a good example. P14 also said they don't have natural data drift problems:
    
    \begin{displayquote}
    The model gets retrained every day, so we don't have the scenario of like: Oh, our models got stale and we need to retrain it because it's starting to make mistakes because data has drifted...fortunately we've never had to deal with [such a] scenario. Sometimes there are bad [training] jobs, but we can always effectively roll back to a different [model].
    \end{displayquote}
    
    However, a few engineers mentioned that natural data shift could cause some hand-curated features and data quality checks to corrupt (P3, P6, P8). P6 discussed a histogram used to make a feature (i.e., converting a real-valued feature to a categorical feature) for an ML model---as data changed over time, the bucket boundaries became useless, resulting in buggy predictions. P8 described how, in their NLP models, the vocabulary of frequently-occurring words changed over time, forcing them to update their preprocessor functions regularly. Our takeaway is that any function that summarizes data---be it cleaning tools, preprocessors, features, or models---needs to be refit regularly.
\end{itemize}

\topic{Takeaway} Unfortunately, none of the participants reported having solved the Goldilocks ML alert problem at their companies. What metrics can be reliably monitored in real-time, and what criteria should trigger alerts to maximize precision and recall when identifying model performance drops? How can these metrics and alerting criteria---functions of naturally-drifting data---automatically tune themselves over time? We envision this to be an opportunity for new data management tools.

\subsubsection{Taming the Long Tail of ML Pipeline Bugs}

In the interviews, we gathered the sentiment that ML debugging is different from debugging during standard software engineering, where one can write test cases to cover the space of potential bugs. But for ML, if one can't categorize bugs effectively because every bug feels unique, how will they prevent future similar failures? Moreover, it's important to fix pipeline bugs as soon as possible to minimize downtime, and {\bf a long tail of possible ML pipeline bugs forces practitioners to have high debugging velocity.}  \blockquote{I just sort of poked around until, at some point, I figured [it] out,} P6 said, describing their ad-hoc approach to debugging. Other participants similarly mentioned that they debug without a systematic framework, which could take them a long time (P5, P8, P10, P18).

While some types of bugs were discussed by multiple participants, such as accidentally flipping labels in classification models (P1, P3, P6, P11) and forgetting to set random seeds (P1, P12, P13), the vast majority of bugs described to us in the interviews were seemingly bespoke and not shared among participants. For example, P8 forgot to drop special characters (e.g., apostrophes) for their language models. P6 found that the imputation value for missing features was once corrupted. P18 mentioned that a feature of unstructured data type (e.g., JSON) had half of the keys' values missing for a \blockquote{long time.} 

\topic{Unpredictable Bugs; Predictable Symptoms} Interestingly, these one-off bugs from the long tail showed similar symptoms of failure. For instance, a symptom of unnatural data drift issues (defined in~\Cref{sec:challenges-pain-dataval}) was a large discrepancy between offline validation accuracy and production accuracy immediately after deployment (P1, P6, P14, P18). The similarity in symptoms highlighted the similarity in methods for isolating bugs; they were almost always some variant of slicing and dicing data for different groups of customers or data points (P2, P6, P11, P14, P17, P19). P14 discussed tracking bugs for different slices of data and only drilling down into their queue of bugs when they observed \blockquote{systematic mistakes for a large number of customers.} P2 did something similar, although they hesitated to call it debugging:

\begin{displayquote}
You can sort of like, look for instances of a particular [underperforming slice] and [debug]---although I'd argue that [it isn't] debugging as much as it is sampling the world for more data...maybe it's not a bug, and [it's] just [that] the model has not seen enough examples of some slice.
\end{displayquote}

\topic{Paranoia Caused by ML Debugging Trauma} After several iterations of chasing bespoke ML-related bugs in production, ML engineers that we interviewed developed a sense of paranoia while evaluating models offline, possibly as a coping mechanism (P1, P2, P6, P15, P17, P19). P1 recounted a bug that was \blockquote{impossible to discover} after a deployment to production:

\begin{displayquote}
ML [bugs] don't get caught by tests or production systems and just silently cause errors [that manifest as] slight reductions in performance. This is why [you] need to be paranoid when you're writing ML code, and then be paranoid when you're coding. I remember one example of a beefy PR with a lot of new additions to data augmentation...but the ground truth data was flipped. If it hadn't been caught in code review, it [would've been] almost impossible to discover. I can think of no mechanism by which we would have found this besides someone just curiously reading the code. [In production], it would have only [slightly] hurt accuracy.
\end{displayquote}

It's possible that many of the bespoke bugs could be ignored if they didn't actually affect model performance. Tying this concept to the data quality issue, maybe all engineers needed to know was when model performance was suffering. But they needed to know \emph{precisely} when models were underperforming, an unsolved question as discussed in~\Cref{sec:challenges-pain-dataval}. When we asked P1, \blockquote{how do you know when the model is not working as well as you expect?}---they gave the following answer:

\begin{displayquote}
Um, yeah, it's really hard. Basically there's no surefire strategy. The closest that I've seen is for people to integrate a very high degree of observability into every part of their pipeline. It starts with having really good raw data, observability, and visualization tools. The ability to query. I've noticed, you know, so much of this [ad-hoc bug exploration] is just---if you make the friction [to debug] lower, people will do it more. So as an organization, you need to make the friction very low for investigating what the data actually looks like, [such as] looking at specific examples.
\end{displayquote}

\topic{Takeaway} Our takeaway is that there is a chicken-and-egg problem in making it easier to tackle the long tail of ML bugs. To group the tail into higher-order categories---i.e., to know what bugs to focus on and what to throw out---we need to know when models are precisely underperforming; then we can map performance drops to bugs. However, to know when models are precisely underperforming, given feedback delays and other data quality assessment challenges as described in~\Cref{sec:challenges-pain-dataval}, we need to be able to identify all the bugs in a pipeline and reason how much each one could plausibly impact performance. Breaking this cycle could be a valuable contribution to the production ML community and help alleviate challenges that stem from the long tail of ML bugs. 

\subsubsection{Multi-Staged Deployments Seemingly Take Forever}

Multiple participants complained that end-to-end experimentation---the conception of an idea to improve ML performance to validating the idea---took too long (P7, P14, P16, P17, P18, P19). {\bf This reveals the synergies between velocity and validating early}: if ideas can be invalidated in earlier stages of deployment, then overall velocity is increased. But sometimes a stage of deployment would take a long time to observe meaningful results---for example, P19 mentioned that at their company, the timeline for trying a new feature idea took over three months:

\begin{displayquote}
I don't have the exact numbers; around 40 or 50\% will make it to initial launch. And then, either because it doesn't pass the legal or privacy or some other complexity, we drop about 50\% of [the launched experiments]. We have to drop a lot of ideas.
\end{displayquote}

The uncertainty of whether projects will be successful stemmed from the unpredictable, real-world nature of the experiments (P18, P19). P19 said that some features don't make sense after a few months, given the nature of how user behaviors change, which would cause an initially good idea to never fully and finally deploy to production:

\begin{displayquote}
You have to look at so many different metrics, and even for very experienced folks doing this process like dozen times, sometimes it's hard to figure out especially when the user's behavior changes very steadily. There's no sudden change [in evaluation metrics] because of one launch; it just usage patterns that change.
\end{displayquote}

P18 offered an anecdote where their company's key product metrics changed in the middle of one of their experiments, causing them to kill a experiment that appeared to be promising (the original metric was improving):

\begin{displayquote}
It was causing a huge gain on the product metrics; it was definitely a green
signal. But as for the product, metrics keep on rotating based on the company's priorities, you know. Is it the revenue at this point?
Is it the total number of, let's say, installs? Or clicks at this particular point of time?
They keep on changing with company's roadmap...
\end{displayquote}

\topic{Takeaway} While most participants were unable to share exact information about the length of the staged deployment process and specific anecdotes about experiments they needed to cancel for privacy reasons, we found it interesting how different organizations had different deployment evaluation practices yet similar pain around failed project ideas due to the highly iterative, experimental nature of ML. We believe there is an opportunity for tooling to streamline ML deployments in this multi-stage pattern, to minimize wasted work and help practitioners predict the end-to-end gains for their ideas.

\subsection{Observed MLOps Anti-Patterns}
\label{sec:challenges-antipatterns}

Here we report a list of MLOps anti-patterns observed in our interviews, or common potentially-problematic behaviors in the ecosystem around ML experiments and deployments.

\subsubsection{Industry-Classroom Mismatch} P1, P5, P7, P11, and P16 each discussed some ML-related bugs they encountered early in their career, right after leaving school, that they knew how to avoid only after on-the-job experience. \blockquote{I learned a lot of data science in school, but none of it was quite like all these things you're [asking],} P7 told us at the end of their interview. P5 said they did a lot of \blockquote{learning by doing.}  P11 provided further insight:

\begin{displayquote}
It was [hard to be], like, thrown into the wild, and have to learn all of this on the job. Coming out of [university in the US with a strong CS program], these are not things that anyone has ever taught right, at least in school...my mindset has always been a little bit more, I guess, practically oriented, even since the academic days, and that's not to say we had great mental models---or frameworks or playbooks---for doing this.
\end{displayquote}

Our interviews with participants didn't focus on what specific skills they could have learned in the classroom that would have prepared them better for their jobs. We leave this to future study and collaborations between academia and industry.

\subsubsection{Keeping GPUs Warm} P5 first mentioned the phrase \blockquote{keeping GPUs warm}---i.e., running as many experiments as possible given computational resources, or making sure all GPUs were utilized at any given point in time:

\begin{displayquote}
One thing that I've noticed is, especially when you have as many resources as [large companies] do, that there's a compulsive need to leverage all the resources that you have.
And just, you know, get all the experiments out there. Come up with a bunch of ideas; run a bunch of stuff. I actually think that's bad. You can be overly concerned with keeping your GPUs warm, [so much] so that you don't actually think deeply about what the highest value experiment is.

I think you can end up saving a lot more time---and obviously GPU cycles, but mostly end-to-end completion time---if you spend more efforts choosing the right experiment to run instead of [spreading yourself] thin. All these different experiments have their own frontier to explore, and all these frontiers have different options.
I basically will only do the most important thing from each project's frontier at a given time, and I found that the net throughput for myself has been much higher.
\end{displayquote}

In executing experiment ideas, we noticed a tradeoff between a guided search and random search. Random searches were more suited to parallelization---e.g., hyperparameter searches or ideas that didn't depend on each other. Although computing infrastructure could support many different experiments in parallel, the cognitive load of managing such experiments was too cumbersome for participants (P5, P10, P18, P19). In other words, {\bf having high velocity means drowning in a sea of versions} of experiments. Rather, participants noted more success when pipelining learnings from one experiment into the next, like a guided search to find the best idea (P5, P10, P18).  P18 described their ideological shift from random search to guided search:

\begin{displayquote}
Previously, I tried to do a lot of parallelization. I used to take, like, 5 ideas and try to run experimentation in parallel, and that definitely not only took my time, but I also focused less. If I focus on one idea, a week at a time, then it boosts my productivity a lot more.
\end{displayquote}

By following a guided search, engineers are, essentially, significantly pruning a tree of experiment ideas without executing them. Contrary to what they were taught in academia, P1 observed that some hyperparameter searches could be pruned early because hyperparameters had such little impact on the end-to-end pipeline:

\begin{displayquote}
I remember one example where the ML team spent all this time making better models, and it was not helping [overall performance]. Then everyone was so frustrated when one person on the controls team just tweaked one parameter [for the non-ML part of the pipeline], and [the end-to-end pipeline] worked so much better. Like we've invested all this infrastructure for hyperparameter tuning a experiment, and I'm like what is this. Why did this happen?
\end{displayquote}

Our takeaway is that while it may seem like there are unlimited computational resources, developer time and energy is the limiting reagent for ML experiments. At the end of the day, experiments are human-validated and deployed. Mature ML engineers know their personal tradeoff between parallelizing disjoint experiment ideas and pipelining ideas that build on top of each other, ultimately yielding successful deployments.

\subsubsection{Retrofitting an Explanation} Right from the first interview, participants discussed uncovering good results from experiments, productionizing changes, and then trying to reason why these changes worked so well (P1, P2, P7, P12). P1 said: 

\begin{displayquote}
    A lot of ML is is like: people will claim to have like principled stances on why they did something and why it works. I think you can have intuitions that are useful and reasonable for why things should be good, but the most defining characteristic of [my most productive colleague] is that he has the highest pace of experimentation out of anyone. He's always running experiments, always trying everything. I think this is relatively common---people just try everything and then backfit some nice-sounding explanation for why it works.
\end{displayquote}

We wondered, why was it even necessary to have an explanation for why something worked? Why not simply accept that, unlike in software, we may not have elegant, principled reasons for successful ML experiments? P2 hypothesized that such retrofitted explanations could guide future experiment ideas over a longer horizon. Alternatively, P7 mentioned that their customers sometimes demanded explanations for certain predictions:

\begin{displayquote}
Do I know why? No idea. I have to convince people that, okay, we try our best.
We try to [compute] correlations. We try to [compute] similarities. Why is it different? I have to make conjectures.
\end{displayquote}

We realized that although they could be false, retrofitted explanations can help with collaboration and business goals. If they satisfy customers and help organize teams around a roadmap of experiment ideas, maybe they are not so bad.


\subsubsection{Undocumented Tribal Knowledge}\label{sec:challenges-antipatterns-tribal} P6, P10, P13, P14, P16, P17, and P19 each discussed pain points related to undocumented knowledge about ML experiments and pipelines amongst collaborators with more experience related to specific pipelines. Across interviews, it seemed like {\bf high velocity created many versions, which made it hard to maintain up-to-date documentation.} P10 mentioned that there were parts of a pipeline that no one touched because it was already running in production, and the principal developer who knew most about it had left the company. P16 said that \blockquote{most of the, like, actual models were trained before [their] time.} P14 described a ``pipeline jungle'' that was difficult to maintain:

\begin{displayquote}
You end up with this pipeline jungle where everything's super entangled, and it's really hard to make changes, because just to make one single change, you have to hold so
much context in your brain. You're trying to think about like, okay this one change is gonna affect this system which affects this [other] system, [which creates]...the pipeline got to the point where it was very difficult to make even simple changes.
\end{displayquote}

While writing down institutional knowledge can be straightforward to do once, P6 discussed that in the ML setting, they learn faster than they can document; moreover, people don't want to read so many different versions of documentation:

\begin{displayquote}
There are people in the team, myself included, that have been on it for several years now, and so there's some institutional knowledge embodied on the team that sometimes gets written down. But you know, even when it does get written down, maybe you will read them, but then, they kind of disappear to the ether. 
\end{displayquote}

Finally, P17 realized that poorly documented pipelines forced them to treat pipelines as black boxes: \blockquote{Some of our models are pretty old and not well documented, so I don't have great expectations for what they should be doing.} Without intuition for how pipelines should perform, practitioner productivity can be stunted.

\vspace{1em}
\topic{Takeaway} The MLOps anti-patterns described in this section reveal that ML engineering, as a field, is changing faster than educational resources can keep up. We see this as opportunities for new resources, such as classroom material (e.g., textbooks, courses) to prescribe the right engineering practices and rigor for the highly experimental discipline that is production ML, and automated documentation assistance for ML pipelines in organizations.   

\subsection{Characterizing the ``MLOps Stack'' for Tool Builders}
\label{sec:challenges-caution}

\begin{figure}
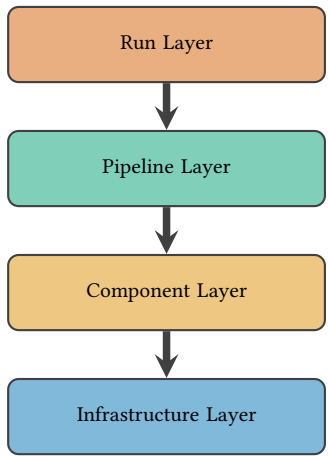

    \centering
         \tikzset{
  every shadow/.style={
    fill=none,
    shadow xshift=0pt,
    shadow yshift=0pt}
}
\tikzset{module/.append style={top color=\col,bottom color=\col}}
    \smartdiagramset{
    set color list={run!50, pipeline!50, component!50, infra!50},
    uniform arrow color=true,
    arrow color=gray!50!black,
    border color=gray!50!black,
    back arrow disabled=true,
    text width=4cm,
    }
    \smartdiagram[flow diagram]{Run Layer, Pipeline Layer, Component Layer, Infrastructure Layer}
    \caption{Layers of tools in the MLOps stack.}
    \label{fig:stackdiagram}
\end{figure}

MLOps tool builders may be interested in an organization of the dozens of tools, libraries, and services MLEs use to run ML and data processing pipelines.
Although multiple MLEs reported having to ``glue'' open-source solutions together and having to build ``homegrown'' infrastructure as part of their work (P1, P2, P5, P6, P10, P12),
an analysis of the various deployments reveals that tools can be grouped into a stack of four layers, depicted in \Cref{fig:stackdiagram} and discussed further in \Cref{app:toolstack}. We discuss the four layers in turn:

\begin{enumerate}
\item \textbf{Run Layer}: A \emph{run} is a record of an execution of an ML or data pipeline (and its components).
Run-level data is often managed by data catalogs, model registries, and training dashboards. \\ 
\textbf{Example Tools}: Weights \& Biases, MLFlow, Hive metastores, AWS Glue

\item \textbf{Pipeline Layer}: Finer-grained than a run, a pipeline further specifies the dependencies between artifacts and details of the corresponding computations. Pipelines can run ad-hoc or on a schedule. Pipelines change less frequently than runs, but more frequently than components. \\ 
\textbf{Example Tools}: Papermill, DBT, Airflow, TensorFlow Extended, Sagemaker

\item \textbf{Component Layer}: A component is an individual node of computation in a pipeline, often a script inside a managed environment. 
Some MLEs reported having an organization-wide ``library of common components'' for pipelines to use, such as feature generation and model training (P2, P6). \\
\textbf{Example Tools}: Python, Spark, PyTorch, TensorFlow

\item \textbf{Infrastructure Layer}: MLEs described a wide range of solutions, but most used cloud storage (e.g., S3), and GPU-backed cloud computing (AWS and GCP). Infrastructure changed far less frequently than other layers in the stack, but each change was more laborious and prone to wide-ranging consequences. \\
\textbf{Example Tools}: Docker, AWS, GCP
\end{enumerate}
We found that MLEs used layers of abstraction (e.g., ``config-based development'') as a way to manage complexity:
most changes (especially high-velocity ones) were minor and limited to the Run Layer, such as selecting hyperparameters. As the stack gets deeper, changes become less frequent: MLEs ran training jobs daily but modified Dockerfiles occasionally. In the past, as MLOps tool builders, we (the authors) have incorrectly assumed uniform user access patterns across all layers of the MLOps stack. Tool builders may want to pay attention to the layer(s) they are addressing and make sure they are not designing tools for the wrong layer(s). 

Additionally, we noticed a high-level pattern in how interviewees discussed the tools they used: engineers seemed to prefer tools that significantly improved their experience with respect to the Three Vs (\Cref{sec:3vs}). For example, experiment tracking tools increased engineers' speed of iterating on feature or modeling ideas (P14, P15)---a velocity virtue. In another example, feature stores (i.e., tables of derived columns for ML models) helped engineers debug models because they could access the relevant historical versions of features used in training such models (P3, P6, P14, P17)---a versioning virtue. MLOps tool builders may want to prioritize ``10x'' better experiences across velocity, validating early, or versioning for their products.



\section{Conclusion}

In this paper, we presented results from a semi-structured interview study of ML engineers spanning different organizations and applications to understand their workflow, best practices, and challenges. We found that successful MLOps practices center around having high velocity, validating as early as possible, and maintaining multiple versions of models for minimal production downtime. We reported on the experimental nature of production ML, aspects of effective model evaluation, and tips to sustain model performance over time. Finally, we discussed MLOps pain points and anti-patterns discovered in our interviews to inspire new MLOps tooling and research ideas. 

\section*{Acknowledgements}

We thank the interviewees for their valuable time and thoughtful responses. We are also grateful to Sarah Catanzaro for connecting us to some of the interviewees, and Alex Tamkin and Preetum Nakkiran for helpful suggestions. We acknowledge support from grants IIS-2129008, IIS-1940759, IIS-1940757, and CNS-1730628 awarded by the National Science Foundation, DOE Grant No. DE-SC0016934, an NDSEG Graduate Fellowship, an NSF Graduate Research Fellowship, funds from the Alfred P. Sloan Foundation, as well as EPIC lab sponsors: Adobe, Microsoft, Google, and Sigma Computing. Work was done while Hellerstein was on leave at Sutter Hill Ventures.

\bibliographystyle{plain}
\bibliography{sample}

\clearpage

\appendix

\begin{table*}[t!]
    \centering
    \noindent\begin{tabular}{|r|p{0.18\linewidth}|p{0.18\linewidth}|p{0.18\linewidth}|p{0.18\linewidth}|}
    \toprule
       & \multicolumn{1}{|c|}{\textbf{Data Collection}} & \multicolumn{1}{|c|}{\textbf{Experimentation}}  &  \multicolumn{1}{|c|}{\textbf{Evaluation and Deployment}} & \multicolumn{1}{|c|}{\textbf{Monitoring and Response}} \\
    \midrule
    \multirow{5}{*}{\textbf{Run}} & \textcolor{purple}{Know what data is available and where it lives} & \textcolor{purple}{Prototype ideas and track results} & \textcolor{purple}{Catch errors in training (e.g., overfitting)} & \textcolor{purple}{Track ML metrics over time} \\
    \cmidrule(lr){2-5}
     & \textcolor{blue}{Data catalogs, Amundsen, AWS Glue, Hive metastores} & \multicolumn{2}{p{0.36\linewidth}|}{\textcolor{blue}{Weights \& Biases, MLFlow, train/test set parameter configs, A/B test tracking tools}} & \textcolor{blue}{Dashboards, SQL, metric functions and window sizes} \\
     \midrule
    \multirow{7}{*}{\textbf{Pipeline}} & \textcolor{purple}{Regularly scheduled, possibly outsourced} & \textcolor{purple}{Ad-hoc or user-triggered, hyperparameter search} & \textcolor{purple}{Scheduled refresh of hold-out validation sets} & \textcolor{purple}{Scheduled computation of metrics and triggered alerts} \\
    \cmidrule(lr){2-5}
    & \textcolor{blue}{In-house or outsourced annotators} & \textcolor{blue}{AutoML} & \textcolor{blue}{Github Actions, Travis CI, Prediction serving tools, Kafka, Flink} & \textcolor{blue}{Prometheus, AWS CloudWatch} \\
    \cmidrule(lr){2-5}
    & \multicolumn{4}{c|}{\textcolor{blue}{Airflow, Kubeflow, Argo, Tensorflow Extended (TFX), Vertex AI, DBT}} \\
     \midrule
    \multirow{8}{*}{\textbf{Component}} &  \textcolor{purple}{Sourcing, labeling, cleaning} &  \textcolor{purple}{Feature generation and selection, model training} & \textcolor{purple}{Running model on hold-out validation set, model compression or rewrite, model serialization} & \textcolor{purple}{Data validation, ML metric computation, tracing predictions} \\
    \cmidrule(lr){2-5}
    & \textcolor{blue}{Data cleaning tools} & \textcolor{blue}{Tensorflow, MLlib, PyTorch, Scikit-learn, XGBoost} & \textcolor{blue}{C++, ONNX, OctoML, TVM, joblib, pickle} & \textcolor{blue}{Scikit-learn metric functions, Great Expectations, Deequ} \\
    \cmidrule(lr){2-5} 
    & \multicolumn{4}{c|}{\textcolor{blue}{Python, Pandas, Spark, SQL}} \\
     \midrule
    \multirow{5}{*}{\textbf{Infrastructure}} & \textcolor{purple}{Velocity} & \textcolor{purple}{Velocity} & \textcolor{purple}{Validate early} & \textcolor{purple}{Versioning} \\
    \cmidrule(lr){2-5}
    & \textcolor{blue}{Annotation schema, cleaning criteria configs} & \textcolor{blue}{Jupyter notebook setups, GPUs} & \textcolor{blue}{Edge devices, CPUs} & \textcolor{blue}{Logging and observability services (e.g., DataDog)} \\
    \cmidrule(lr){2-5}
    & \multicolumn{4}{c|}{\textcolor{blue}{Cloud (e.g., AWS, GCP), compute clusters, storage (e.g., AWS S3, Snowflake), Docker, Kubernetes}} \\
     \bottomrule
    \end{tabular}%
    \caption{Primary \textcolor{purple}{goals} and \textcolor{blue}{tools} for each layer in the MLOps stack and routine task in the ML engineering workflow. }
    \label{tab:stack}
\end{table*}

\section{Semi-Structured Interview Questions}
\label{app:interviewq}

In the beginning of each interview, we explained the purpose of the interview---to better understand processes within the organization for validating changes made to production ML models, ideally through stories of ML deployments. 
We then kickstarted the information-gathering process with a question to build rapport with the interviewee, 
such as \emph{tell us about a memorable previous ML model deployment}. 
This question helped us isolate an ML pipeline or product to discuss. We then asked a series of open-ended questions:

\begin{enumerate}
    \item \textbf{Nature of ML task} 
    \begin{itemize}
        \item What is the ML task you are trying to solve? 
        \item Is it a classification or regression task? 
        \item Are the class representations balanced? 
    \end{itemize}
    \item \textbf{Modeling and experimentation ideas} 
    \begin{itemize}
        \item How do you come up with experiment ideas? 
        \item  What models do you use? 
        \item How do you know if an experiment idea is good?
        \item  What fraction of your experiment ideas are good?
    \end{itemize}
    \item \textbf{Transition from development to production}
    \begin{itemize}
        \item What processes do you follow for promoting a model from the development phase to production?
        \item How many pull requests do you make or review?
        \item What do you look for in code reviews?
        \item What automated tests run at this time?
    \end{itemize}
    \item \textbf{Validation datasets} 
    \begin{itemize}
        \item How did you come up with the dataset to evaluate the model on?
        \item Do the validation datasets ever change?
        \item Does every engineer working on this ML task use the same validation datasets?
    \end{itemize}
    \item \textbf{Monitoring} 
    \begin{itemize}
        \item Do you track the performance of your model?
        \item If so, when and how do you refresh the metrics?
        \item What information do you log? 
        \item Do you record provenance?
        \item How do you learn of an ML-related bug?
    \end{itemize}
    
    \item \textbf{Response} 
    \begin{itemize}
        \item What historical records (e.g., training code, training set) do you inspect in the debugging process? 
        \item What organizational processes do you have for responding to ML-related bugs?
        \item Do you make tickets (e.g., Jira) for these bugs?
        \item How do you react to these bugs?
        \item When do you decide to retrain the model?
    \end{itemize}
\end{enumerate}

\section{Interview Transcripts}
\label{app:transcripts}

\begin{figure*}
    \centering
    \begin{subfigure}[b]{0.48\linewidth}
    \centering
    \resizebox{\linewidth}{!}{\input{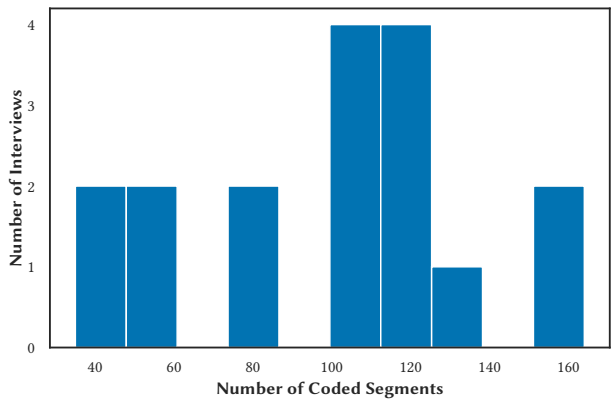}}
    \caption{Histogram of number of coded segments in each interview.}
    \label{fig:transcripthistograms-codes}
    \end{subfigure}
    \begin{subfigure}[b]{0.48\linewidth}
    \centering
    \resizebox{\linewidth}{!}{\input{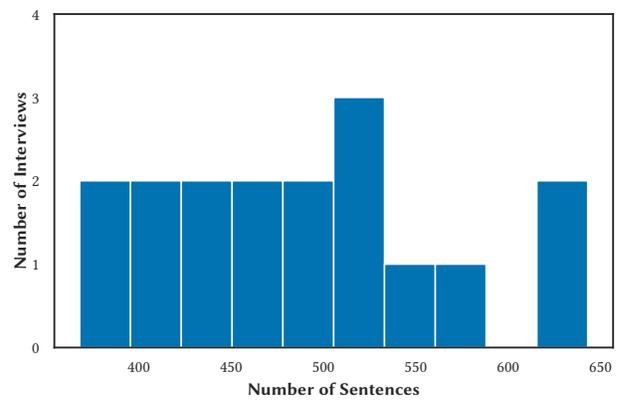}}
    \caption{Histogram of number of sentences in each interview.}
    \label{fig:transcripthistograms-sentences}
    \end{subfigure}
    \caption{Interview transcript statistics. Each histogram has 10 equally-spaced buckets.}
    \label{fig:transcripthistograms}
\end{figure*}

Histograms of the number of codes and sentences in the interview transcripts are shown in \Cref{fig:transcripthistograms-codes,fig:transcripthistograms-sentences}, respectively.

\section{Codes}
\label{app:codes}

\begin{table*}[t!]
    \centering
    \resizebox{\linewidth}{!}{
\begin{tabular}{lllrr}
\toprule
{} &                  \textbf{Parent Code} &                                         \textbf{Code} &  \textbf{\# Coded Segments} &  \textbf{\# Transcripts} \\
\midrule
1  &                 known challenges &                        data drift/shift/skew &                         15 &         10 \\
2  &      monitoring and response (+) &                              live monitoring &                         21 &          9 \\
3  &                       Python (+) &                                      Jupyter &                         16 &          8 \\
4  &        evaluation and deployment &                     build the infrastructure &                         16 &          8 \\
5  &                    data pipeline &                  data iteration,  fresh data &                         15 &          8 \\
6  &                    fast \& simple &  high iteration speed,  agile,  rapid cycles &                         24 &          7 \\
7  &                  production bugs &                           debugging and bugs &                         18 &          7 \\
8  &                            tests &                                   AB Testing &                         14 &          7 \\
9  &             software development &                                 pull request &                         13 &          7 \\
10 &                    data pipeline &                       pipeline on a schedule &                         13 &          7 \\
11 &                       operations &                  model training \& retraining &                         12 &          7 \\
12 &                      data ingest &                      automated featurization &                          9 &          7 \\
13 &        evaluation and deployment &                                        CI/CD &                          8 &          7 \\
14 &                           trends &        per-customer model and many customers &                         15 &          6 \\
15 &                 known challenges &                               feedback delay &                         13 &          6 \\
16 &        evaluation and deployment &                       metrics and validation &                         12 &          6 \\
17 &       metrics and validation (+) &                                     accuracy &                         11 &          6 \\
18 &                           models &                                deep learning &                          9 &          6 \\
19 &                       sandboxing &               offline demonstration of value &                          7 &          6 \\
20 &                 apps \& use-cases &                                      ranking &                          7 &          6 \\
\bottomrule
\end{tabular}
}
    \caption{Top 20 codes, ordered by the number of distinct transcripts the codes were mentioned in, descending.}
    \label{tab:freqcodes}
\end{table*}

\begin{sidewaysfigure*}[h!]
    \vspace{12cm}
    \hspace{-3cm}
    \includegraphics[scale=1.5]{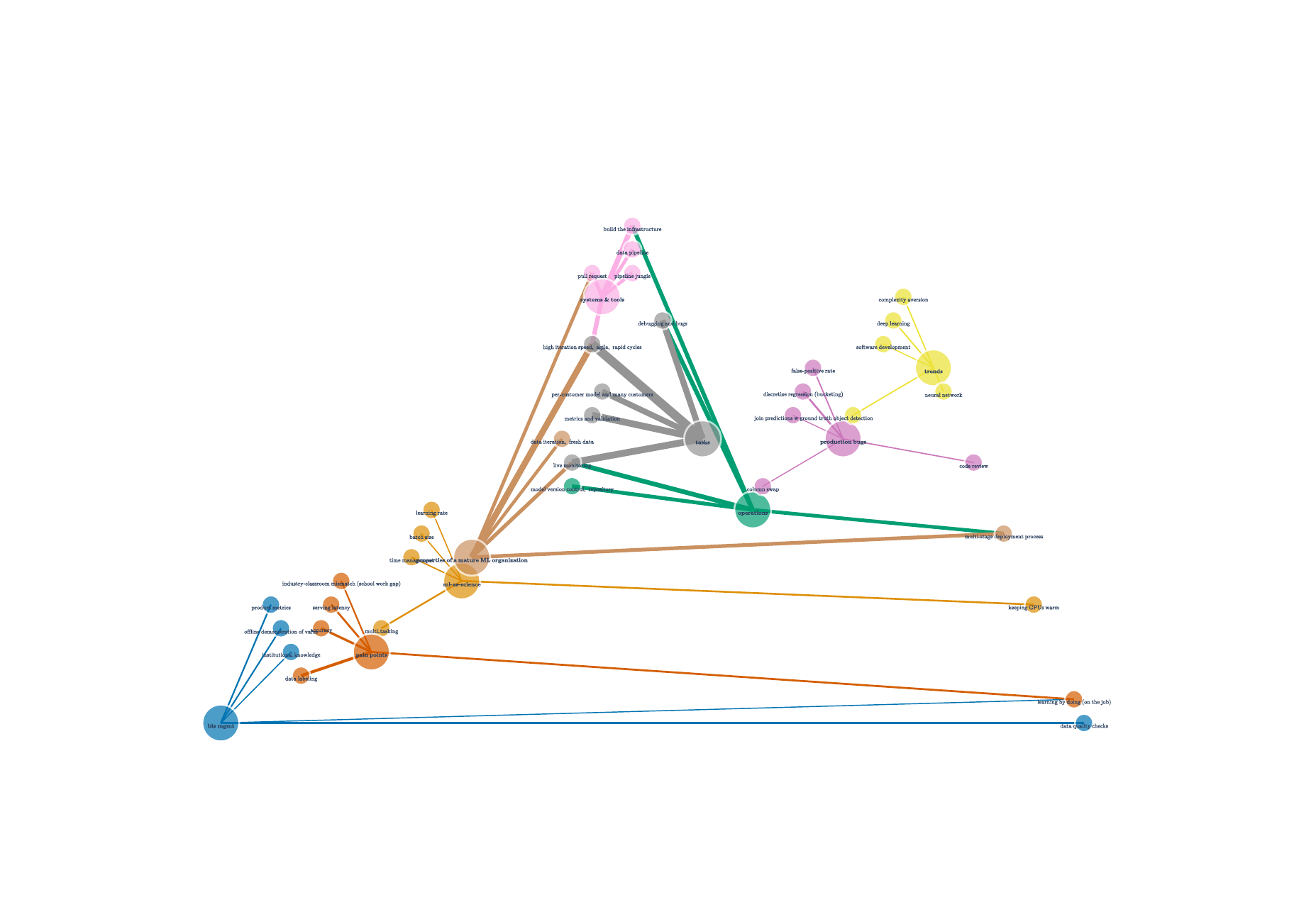}
    \caption{Correlated codes for each top-level code. Each edge is weighted by the occurrence count for its pair of codes.}
    \label{fig:correlatedcodes}
\end{sidewaysfigure*}

Across the interview transcripts, we had a total of 1766 coded segments, with exactly 600 unique codes. We organized codes into hierarchies. \Cref{tab:freqcodes} shows the most frequently occurring codes, ordered by the number of distinct interviews the codes appeared in (not the raw number of occurrences across all documents). \Cref{fig:correlatedcodes} displays the top five correlated codes for each top-level or parent code. Two codes are correlated if they occur within twenty sentences of each other.

\section{MLOps Tool Stack}
\label{app:toolstack}

\Cref{tab:stack} shows common tools used by MLEs across layers of the stack and tasks in the production ML lifecycle.

\end{document}